# Testing for the Continuous Spectrum of X-Rays Predicted to Accompany the Photoejection of an Atomic Inner Shell Electron

Philip Jacobson,[1] Andrija Rasovic,[1] Arthur Campello,[1] Chase Goddard,[1] Matthew Dykes,[1] Yuchao Chen,[1]
J.Y. Peter Ko,[2] Stanislav Stoupin,[2] Gwen Gardner,[1] Justin Oh,[1] and Carl Franck[1]

*[1]Laboratory of Atomic and Solid State Physics, Cornell University, Ithaca, NY, USA*
*[2]Cornell High Energy Synchrotron Source (CHESS)*

Echoing classical physics, quantum electrodynamics predicts the release of a spectral continuum of electromagnetic radiation upon the sudden acceleration of charged particles in quantum matter. Despite apparent theoretical success in describing sister nuclear processes, known as internal bremsstrahlung, following nuclear beta decay and K capture, the situation of the photoejection of an electron from an inner shell of an atom, intraatomic bremsstrahlung (IAB), is far from settled. In this paper we present fresh measurements which rely on contemporary signal processing as well as the high flux available from a synchrotron radiation source to revisit the problem by photoejecting electrons from the innermost shell of copper. For the first time we have sufficient sample statistics to measure the expected spectra at the level expected by contemporary theory. Furthermore, we employ sufficiently thin targets to overcome secondary scattering artifacts. Our approach applies the fluorescence coincidence method to guard against extraneous scattering and multiple incident photon processes. Our observations set a severe upper limit on the rate for IAB: We conclude that current theory overpredicts, by at least 5 sigma, the measured rate for K shell IAB in copper in the range of detected energies below the K fluorescence energy.

## I.    INTRODUCTION

The infrared divergence (IRD) encountered in charged particle processes wherein an outgoing electron radiates a soft photon has played a critical role [1] in the quantum theory of charged particle-photon interactions (Quantum Electrodynamics, QED). Such IRDs in amplitudes have been tamed into cancellation in the computation of the scattering cross sections of charged leptons. However, as emphasized by Pratt[2] and first pointed out by Gavrila[3], the ejection of a bound electron from an atom by photoabsorption of an incident photon is expected to reveal a continuous spectrum that diverges in spectral intensity as the energy of the accompanying radiated photon vanishes. Thus, the IRD of QED should be directly detectable in the process we refer to as intraatomic bremsstrahlung (IAB). The aim of this study is to observe such radiation in the form of x-rays. Following the arguments in Refs. [4] and [5], the process can be understood semiclassically by assuming that the ejection process is sufficiently rapid. Eqn. (6) of Ref. [5] sets the very conservative (see Ref. [6]) upper limit for such behavior on the scattered photon energy $E_S$ to be $E_{div}$ as in:

$$E_S < E_{div} \equiv 2^{\frac{1}{2}} \frac{hc}{a_0} \left(\frac{T_0}{mc^2}\right)^{\frac{1}{2}}$$
$$= 33 \ keV \ (\tfrac{T_0}{mc^2})^{1/2} \qquad (1)$$

where $h$ is Planck's constant, $a_0$ is the Bohr radius, $T_0 = (E_0 - E_B)$ is the kinetic energy of the photoejected electron with binding energy $E_B$, $E_0$ being the energy of an incident photon, and $m$ is the electron rest mass. We take these photons to be linearly polarized. In this limit, the emitted radiation expected when a charged particle is accelerated from rest to a high velocity is given by [5], [7]:

$$\frac{d^2\sigma_{IAB}}{dE_S d\Omega_S} = \frac{\alpha}{4\pi^2 E_S} \frac{2(E_0 - E_B)}{mc^2} \left(\frac{4}{5} - \frac{2}{5}cos^2\theta\right)\sigma_K(E_0) \quad (2)$$

where $\theta$ is the angle between the direction of the scattered radiation and the polarization vector of the incident radiation and $\sigma_K$ is the K-shell photoabsorption cross section at energy $E_0$. Later, Bergstrom, Pisk, Suric and Pratt reinforced this prediction, called the Low Energy Theorem (LET), with contemporary QED theory [8].

To summarize our results: with high probability we did not observe the predicted radiation. In earlier attempts, the situation was quite the opposite: in 1989 and 1990, resonance-enhanced experiments by Briand et al. [9, 10] observed rates as much as a factor of 15 higher, as shown in Fig. 33 of Ref. [8], than predicted. A similar discrepancy had been reported in 1977 in observations employing a fluorescence coincidence detection method that isolated K shell excitations [11]. Prior to the work of Briand, Marchetti and Franck used fluorescence coincidence with a thick target to demonstrate how the IAB signal could be masked by secondary bremsstrahlung production by photoelectrons and argued that this was the case in earlier experiments which claimed an observation of IAB [5,12]. Later experiments also confirmed the contribution of secondary effects in obscuring the predicted IAB signal in fluorescence coincidence spectroscopy [13, 14]. Similarly, Bergstrom *et al.* [8] pointed out that the Low Energy Theorem's prediction for the problem of the photoejection of an electron from an atomic system has not been validated experimentally. Rather the observational situation has long been unsettled. Bergstrom and Pratt [15] took the position in 1997 that "While the theoretical situation here for the infrared features seems clear, the experimental situation is not." In contrast, sister nuclear experiments going back to at least 1962 on beta emission and electron capture have found agreement with predictions. For example, the infrared divergences that are expected to accompany beta decay [16] and electron K capture [17] have been closely reproduced in experiment. [16, 17, 18, 19] In this paper, we return to the atomic problem, exploiting a high flux x-ray source, contemporary x-ray detection processing techniques, ultrathin film copper targets and fluorescence coincidence detection to suppress extraneous background scattering in order to test for intraatomic bremsstrahlung experimentally. We validate our methodology in a sufficiently thick sample through the detection at the expected rate of ordinary bremsstrahlung due to K shell photoelectrons. From our estimate for this secondary process, we conclude that for our thinnest specimens, the contribution to the scattered photon signal is insignificant. This yields the central result of this paper: that the contemporary theory for intraatomic bremsstrahlung is ruled out by placing a tight upper limit on the measured rate from scattered photon production in the low energy range below K fluorescence.

## II. EXPERIMENTAL METHODS INCLUDING CONTROL EXPERIMENTS

### A: Improvements Over Earlier Fluorescence Coincidence Measurements of Inelastic Scattering

Our approach is a follow-up to our earlier efforts [20, 5, 21] that demonstrated that for a highly time- structured and fluctuating source provided by a synchrotron, a properly executed fluorescence coincidence method was a reliable tool for investigating low cross section inelastic scattering from electrons in a particular shell. This was accomplished by suppressing backgrounds due to multiple incident photons and extraneous scattering due to random scattering in the hutch. In that work a successful match between theory based on Hartree-Fock-Slater wavefunctions and experiment was provided for x-ray Raman scattering through the intermediate momentum transfer regime where spectral overlap with inelastic scattering for more loosely bound electrons obscured the signal from nonresonant inelastic x-ray Raman scattering by K shell electrons [20]. This conclusion was subsequently supported by an S-matrix calculation by Bergstrom *et al.* [8]. We demonstrated that an attempt to observe IAB was thwarted by the secondary scattering [5] within the too-thick target.

For the current experiment we returned to the upgraded version of the Cornell High Energy Synchrotron Source (CHESS), using a 46 keV undulator beam incident on thin Cu films. With $E_B$ = 9.0 keV [22].this gives $E_{div}$ in Eqn. (1) equal to 9 keV so Eqn. (1) is satisfied and Eqn. (2), the LET prediction, is expected to hold. The source bandwidth was reduced to approximately 50 eV by using a single-reflection diamond monochromator. Higher harmonics were rejected with a rhodium mirror. The incident beam had horizontal linear polarization. Heavy metal slits were used to define the incident beam spot on target to have 0.5 mm extent both horizontally and vertically and an aluminum wheel with a variety of thicknesses was used to control the incident flux. He and Ar ionization chambers were used to measure the strength of the incident beam. Throughout the experiment, the beam flux was maintained at approximately $10^{10}$ photons s$^{-1}$. To control secondary scattering we employed ultrathin specimens. To reduce scattering from air, the experiment was fully contained within the He atmosphere of the scattering chamber, shown in Figures 1a and 1b. The incident beam energy was chosen to minimize the ratio of secondary emission to the predicted IAB emission (see later discussion of secondary processes) since it was predicted to be far greater using a lower energy incident beam option of 20 keV. For detection, we used Hitachi Vortex 90 EX and ME4 silicon drift detectors, [23] which are one-element and four-element detectors respectively as shown in Fig. 1a. They were outfitted with specially-designed steel snouts (Figure 1b) to eliminate concerns of crosstalk between them (see later discussion on how this requirement was relaxed). The one and four element detectors both measured scattered and fluorescence radiation in the horizontal plane in directions approximately 30° and 45° from the forward scattering direction, respectively.

The Cu films used in the experiment were of thicknesses 40, 80, 160 and 320 nm, considerably thinner than the 7.7 micron thick foils used in our earlier work. [5] These thin films were prepared by thermal evaporation through 5 mm diameter contact masks onto cleaved single crystals of sodium chloride. Water polishing of the substrates provided smooth surfaces that resulted in sufficiently strong films. The films were floated off the substrates and onto deburred sheet plastic target holders of diameter 2, 3 or 4 mm that had been waterjet cut so that all samples could be positioned without adjustment in the precision cut target holder mount shown in Fig. 1b. Residual water was removed by evaporation from the vertically-held target holder in a desiccator. With practice capturing and transferring films this yielded mirror-finish targets, likely under tension. The thinnest films were remarkably strong with one surviving a day of measurement which included multiple mountings and unmountings in the scattering chamber. While most films used remained completely intact, minor tears occasionally appeared.

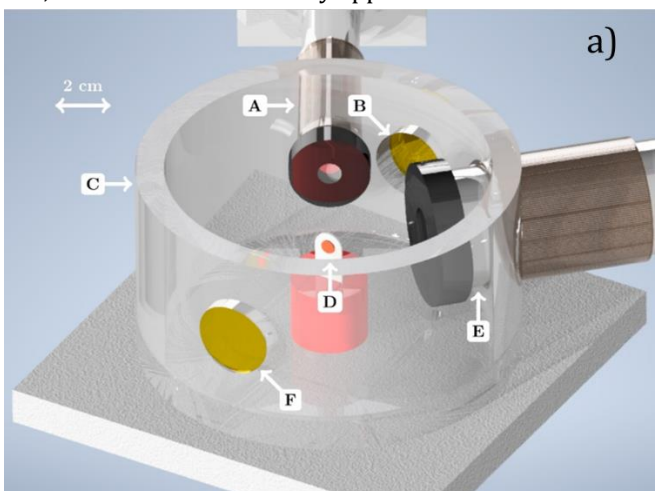

FIG. 1a: Schematic of the Scattering Chamber. A is the one element detector, B is the Kapton film covered main beam exit port, C is the helium (1 Atm.) filled chamber (input and output helium supply lines and chamber cover not shown), D is the target mount, E is the four-element detector, F is the Kapton film covered incident beam port.

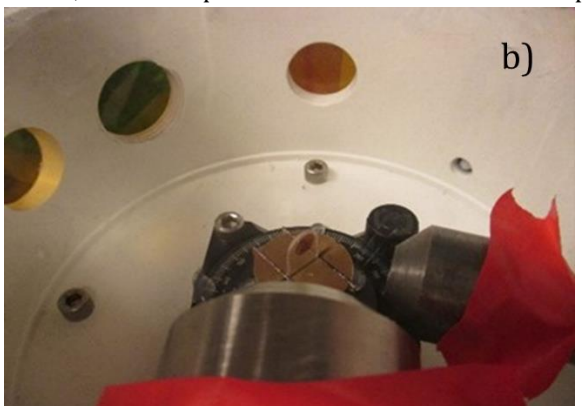

FIG. 1b: Inside View of the Scattering Chamber. Ultrathin copper samples are placed in the middle of the chamber on plastic sample holders.

In order to assess the strength of Compton and elastic scattering, we examined a wide range of scattered photon energies, up to and including the incident energy at 46 keV, although the detector efficiency was only 5% at this energy, compared to approximately 100% at the fluorescence energy. The resultant elastic spectra were noticeable even for an empty target holder, indicating the importance of possible stray scattering, for example from the detector snouts. For an 80 nm thick target this signal depended on the type of detector (one element vs. four element Vortex) employed and at its peak was typically (when corrected for detector efficiency) comparable to the peak rate for K fluorescence.

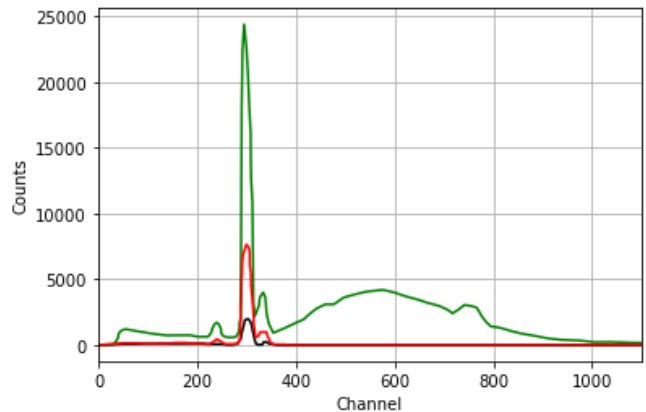

FIG. 2: Pileup Anomaly Demonstrated. In this measurement of the K fluorescence lines (channel number proportional to photon energy, absolute number of counts on the ordinate) for a 320 nm target, equal duration of exposure but with variable attenuation as indicated by color green : red : blue = 432 :99 :26 (relative incident flux values from upstream gas ionization chamber) corresponding to the greatest, intermediate and least detected count spectrum respectively. For the highest incident flux, we see a distorted spectrum in the form of a broad bump at roughly twice the fluorescence line energy indicating pileup. We avoided this distortion with adequate attenuation.

At sufficiently high flux, we noticed that the expected $K_\alpha$ and $K_\beta$ fluorescence spectra around 8 keV were distorted and were surprised that it was not a simple case of seeing a double pulse line at about 16 keV, rather a broad feature at twice the fluorescence energy, as shown in Figure 2. We therefore always attenuated the beam for each different specimen thickness and ran with the least attenuation needed to provide the same fluorescence spectra at lower flux levels. This precaution addresses concern over the detectors and subsequent signal processor being overwhelmed by high instantaneous event rates for our low duty cycle source.

Surprisingly, extraneous signals were detected when incident photons were not on the sample. The evidence is that despite our source's 6% duty cycle, a high speed digital oscilloscope connected to our detectors observed signals synced to the orbits of the positron [24] trains circulating in the storage ring, approximately 40% of the time. We speculate that this might be attributed to stray pickup along the approximately 10 m long cables connecting the detectors to the oscilloscope. For the data presented here, short cables were used.

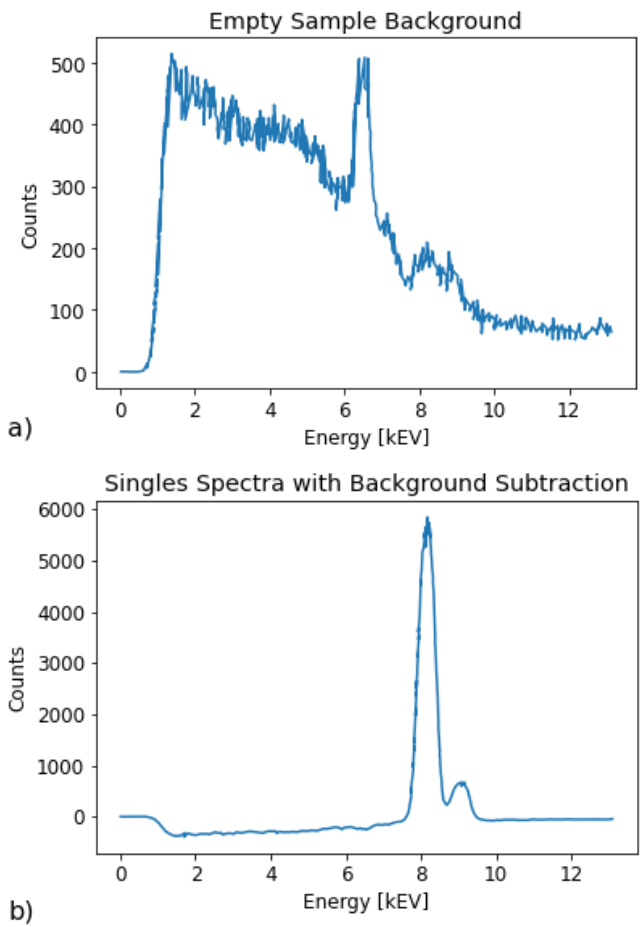

FIG. 3: a) A singles spectra from a run on an empty sample holder. b) Singles spectra for a 80 nm film after background subtraction using the empty target spectra demonstrating a negative signal in the energy range of interest (3-7 keV). We conclude that testing for IAB in this manner is a hopeless task.

## B.    Singles Spectra

In considering whether coincidence detection was necessary for the experiment, several raw spectra without coincidence detection, referred to as singles spectra, were examined for both empty sample holders as well as with the Cu films. The empty sample holder background was significant at low energies, as shown in Figure 3a. When it was subtracted after normalization for the number of incident photons from singles spectra taken with a target, as shown in Figure 3b, the results were not physical: negative results were found for the low energies in the spectral region of interest. We suspect that background scattering from wings of the incident beam may have caused this effect. We concluded that the singles approach for testing for the IAB signal is not viable. In addition, we used Eqn. (2) to compare the measured singles fluorescence strength with the strength expected for the IAB signal at 3 keV, which is an optimum energy to probe after taking into account the expected rise in the IAB signal with reduced scattered energy versus falloff due to diminishing detector sensitivities at low energy. Even without taking into account the additional attenuation and falloff in detector sensitivity, we found that the expected rate was between a factor of 400 and 10,000 below the measured rate for an 80 nm sample. This implied that it was a hopeless task to find the IAB signal within the singles spectra. This provides further motivation for the use of the fluorescence coincidence technique.

## C.    The Coincidence Technique

Recalling our earlier studies described at the start of this section, the application of fluorescence coincidence detection was not only a means of narrowing our focus to excitation of K shell electrons to the exclusion of all others, but also an effective means of suppressing backgrounds. For example, in the pursuit of weak signal scattering, e.g. nonresonant inelastic scattering spectroscopy, this approach affords a means of discrimination against detection events that are due to collisions of incoming photons with unintended targets such as atmospheric scattering and scattering from x-ray optical elements such as shielding.[20, 21]

We now consider the coincidence rate of scattered ($S$) and fluorescence ($F$) photons. Since the excitation and decay processes, although so close in time as to be effectively simultaneous, are separable, the measured triple differential cross section for coincidence detection is given by

$$\frac{d^3\sigma_{IAB}}{dE_S d\Omega_S d\Omega_F} = \frac{\eta_K}{4\pi} \frac{d^2\sigma_{IAB}}{dE_S d\Omega_S} \qquad (3)$$

where $\eta_K$, the fluorescence probability for the decay of a K vacancy, is 0.44 according to Ref. [25].

An XIA xMap [26] digital pulse processor was used to perform all necessary signal processing for real-time data acquisition. Data for coincidence detection in post processing was recorded using the xMap's event mode, also known as list mode, which time-stamps recorded events independently of the synchrotron's positron bunch arrival timing signal. We note that checks for detector pileup during the run were performed by operating the xMap without time stamping. To calculate events in coincidence, data from each of the xMap's four channels were sorted into 20 ns time bins, which is the maximum resolution allowed by the xMap. One channel was arbitrarily designated the fluorescence channel, in which events were restricted to those in the $K_\alpha$ and $K_\beta$ fluorescence region. Events in the remaining channels (we could only use three of the four elements in the ME4 detector since the xMap had only four input channels), designated the scattering channels, that fell in the same time bin as a fluorescence photon were designated a coincidence event. In our final data reduction, we rotated the arbitrary "fluorescence channel" label amongst each of the four channels in order to boost our statistics. As is discussed below, we statistically tested that this bypassing of our anti-cross-talk shields was valid. Accidental coincidence event rates resulting from two uncorrelated incident photons were calculated by introducing a time-offset between the detection events in the scattered photon and fluorescence detection channels and checking for coincidence. Enhanced precision on the accidental event rate could be achieved through averaging over many time-offsets. We emphasize that we were able to accomplish this because of the time-stamped data recorded by the xMap. In contrast, the fast-slow coincidence detection scheme traditionally used [21] in situations where accidental rates are significant requires that half of beam time is devoted to measuring delayed coincidence. Furthermore, such schemes do not allow arbitrarily great precision in establishing the accidental rates since only a single fixed delay can be employed. In contrast, our scheme did not sacrifice beam time since accidental and in-coincidence events were acquired at the same time. We note that the time structure of the stored beam in the ring is highly nonuniform (and has a low duty cycle (6%)): three positron trains of roughly 50 ns duration and spaced roughly 200 ns apart orbit every 2500 ns.

Due to the highly structured timing nature of the source, it was necessary that the time-offset used to compute the accidental rate be an integer multiple of the storage ring's orbital period. As shown in Figure 4, an offset by an integer multiple of periods is necessary to avoid underestimating the accidental rate. By comparison, any time offset greater than the time for the atomic process of excitation and subsequent decay would be suitable with a Poissonian source.

As discussed in Ref. [21], note that the uncertainty due to orbit fluctuations can in principle matter. In that earlier work, we switched between in coincidence and delayed coincidence to measure accidentals every 4 seconds. In principle shorter term fluctuations in the beam would have created excess noise in the accidental rate correction. In the present work, by pooling 10 orbits of data to provide the accidental rate measurement, we not only significantly lowered the accidental rate measurement uncertainty from that provided by a single orbital period offset as used in our earlier work but we decreased the characteristic time that matters for orbital fluctuations by a factor of $1.6 \times 10^5$.

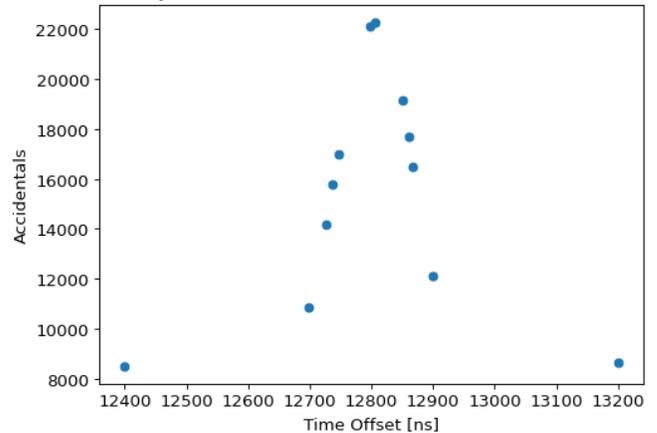

FIG. 4: Accidental coincidence rate versus time offset. Note that an integer multiple (in this case five) of the synchrotron period, corresponding to approximately 12800 ns, provides the largest value for the accidental rate.

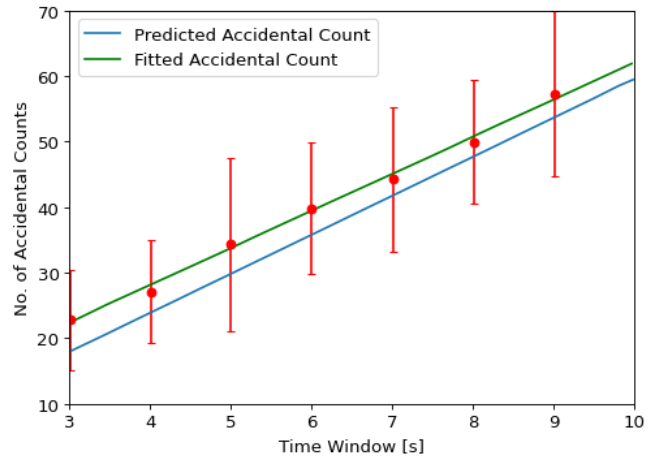

FIG. 5: Accidental Counts: Measured (Red with Green (upper) Fitted Line) vs. Predicted (Blue (lower) Line) as a Function of the Size of Coincidence Time Widow, Using a Radioactive Source.

Prior to the synchrotron run reported here, to test our ability to properly perform coincidence detection with an xMap, we employed an Am 241 radioactive source providing 60 keV photons to excite a copper target and monitored the resultant photons (scattered and fluorescence) using a Vortex detector and a Canberra germanium detector[27]. Realizing that for such a source, in contrast to a synchrotron source, the flux $I$ is steady, the accidental rate can be estimated from the singles rate for the accidental rate given by Eqn. 1a of Ref. [21] the time averages obey the following condition.

$$\langle I^2 \rangle = \langle I \rangle^2 \qquad (4)$$

Since the 0.9 microCi source employed was so weak, a very wide coincidence window of at least four seconds was required to accumulate a significant accidental counts. The comparison between the measured and predicted accidental rate is excellent, as shown for a wide range of time windows in Figure 5.

Turning from the radioactive source to our observations with the time-structured synchrotron radiation source, we again ask to what extent the measured singles rate can be used to estimate the measured accidental rate. This provides insight into the degree to which the fluctuations in the beam intensity causes the condition Eqn. (4) to fail and demonstrates the power of the fluorescence coincidence method in the face of such fluctuations. To do this we randomly selected eight of the 19 runs performed to compute U, the ratio of the measured to the predicted accidental rate. The results are given in the histogram in Figure 6. We see that U varies over a range of factor of 4 centered on the value 1, which is the result expected for a Poissonian source for which Eqn. (4) is obeyed. We conclude that while ignoring beam fluctuations produced a good rough estimate, it is essential that the accidental rate be measured to provide an accurate value for our synchrotron-based measurements.

Recall that we applied attenuation to prevent distortion of the fluorescence spectrum. We note that the resultant event rate of less than about 5 kHz is well below the xMap's maximum event processing rate of 1 MHz, even accounting for the small duty cycle of the source.

### D. Normalization

As an added check on the experiment, we predicted the fluorescence rate for various target thicknesses using

$$n_F = \sigma_K \frac{\eta_K}{4\pi} \frac{\rho N_A}{M} \Phi \, \Omega_F \, t \qquad (5)$$

where $n_F$ is the fluorescence rate, $\sigma_K$ is the Cu K-shell photoabsorption cross section, $\rho$ is the mass density, $N_A$ is Avogadro's number, $M$ is the molar atomic mass, $\Omega_F$ is the solid angle subtended by the fluorescence detector, $\Phi$ is the beam flux, and $t$ is the target film thickness. These predictions were generally several times in excess of the measured results, as shown in Table I. In order to address this discrepancy, we normalized the measured signals by dividing by the measured fluorescence rate in consistency with Eqn. (3).

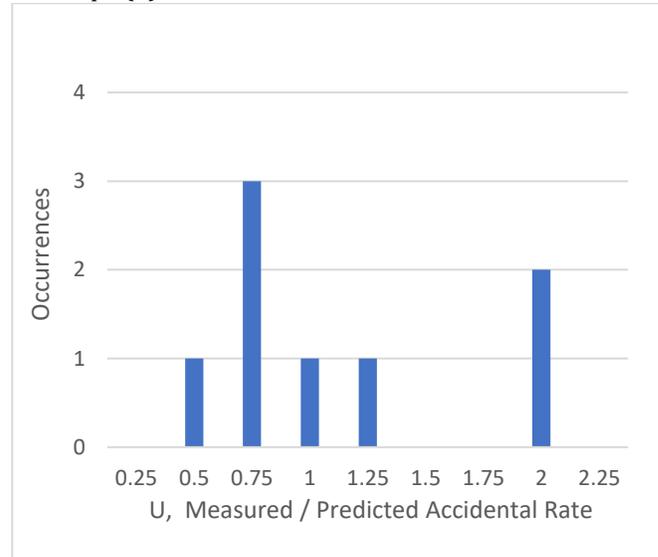

FIG. 6: Histogram of U, Ratio of the Measured to the Predicted Accidental Rates

This normalization greatly simplifies comparison of our experiment with the predicted rate expressed in Eqn. (2). In particular the target thickness, incident flux and the photoabsorption cross section are divided out.

TABLE I: Ratios of predicted to measured fluorescence rates for selected exposures across various sample thicknesses. We address concerns over this discrepancy for the coincidence signal of interest by appropriate normalization according to the measured fluorescence rate.

| Sample Thickness (nm) | Predicted: Measured Fluorescence Rates |
|---|---|
| 320 | 5.97 |
| 160 | 9.31 |
| 80 | 4.58 |
| 80 | 4.55 |
| 40 | 2.97 |
| 40 | 2.80 |

### III. DATA REDUCTION AND ANALYSIS

Before we recognized that we could exchange the roles of the four detection elements, we declared the single element detector to be the fluorescence detector and the three working channels from the four element detector to be the scattered photon detectors. This was primarily done to prevent the so-called cross-talk that occurs when photons scatter between two close detectors. To confirm that cross-talk was negligible, we constructed three other coincidence spectra from the data from our 40 nm samples, each time assigning one of the other three channels of our silicon drift detector as the fluorescence channel. Then, with a t-test, we compared these three sets of data to the original results we obtained by declaring the single element detector as the fluorescence channel. We found no significant difference with an uncertainty that ranged between 3% and 45% and concluded that cross-talk is unimportant. Thus, we were able to significantly increase our coincidence statistics.

To read data from the xMap, we used code provided by XIA. The coincidence calculation was performed using custom code written in Python. To efficiently compare fluorescence events with events in the scattered channel, we stored all fluorescence events in a hash map to facilitate fast comparison with each scattered event. We calculated the number of coincidence counts per 0.5 keV energy bins between the two data samples (designated fluorescence detector and the designated scattered photon detectors). Then, we offset one of the data sets by an integer number of storage ring orbital periods and repeat the process to calculate accidental coincidences. Our final accidental coincidence rate was the average of all these accidentals over ten of those shifts (which means by Poisson statistics that the uncertainty in the accidental rate deduced in this manner is only about 32% of the uncertainty in the accidental rate portion of the in-coincidence measured rate). We used a 100-core computing cluster to perform our calculation of excess coincidence rate. This took a few days since our data, acquired over several days of beamtime, totaled over a terabyte.

## A.    Corrections

The data was corrected for the quantum efficiency of detectors and sources of attenuation including self-absorption in the target, the helium atmosphere and Kapton tape over the detectors' windows. The largest correction comes from the Kapton tape, which attenuates the signal by as much as 80% at the low end of the energy range of interest (2.5 keV and above). The corrections for self-absorption in the target were made using the Beer Lambert Law for radiation exiting at 45° to the incident. As we will see the fact that one of the detectors was at a 30° scattering angle provides an insignificant difference.

## B.    Prediction of Secondary Processes

As emphasized in Ref. [5], secondary processes in which a pair of photons from different atoms are produced at the same time are indistinguishable from genuine IAB with fluorescence events. Thanks to the high flux provided at our source, we could minimize such effects by reducing sample thickness and comparing results as a function of sample thickness. In order to provide evidence for the role of secondary processes, we applied the results of Ref. [5] for the most important secondary process: photoelectrons generated uniformly throughout the sample by absorption of incident x-rays followed by bremsstrahlung radiation by these same electrons elsewhere in the target. This emission is averaged over all positions in the sample where photoelectrons could be created and directions of photoelectrons exiting the atoms. It is shown in Ref. [5] that the differential probability of bremsstrahlung emission $P_B$ from collisions with nuclei of atoms with atomic number $Z$ by the primary photoelectrons is given by:

$$\frac{d^2 P_B}{dE_S d\Omega_S} = \frac{2kZ}{4\pi} \frac{T_0 - E_S}{E_S} \qquad (6)$$

where $k$ is a function of thickness $t$ given by:

$$k(t) = (1.4 \cdot 10^{-6} \ keV^{-1})(\frac{t}{D})^{0.75} \qquad (7)$$

where $D$ is the electron range for the photoelectron in this process which in copper following Ref [5] is 4600 nm. using the continuous slowing down approximation from Ref. [28]. Again, following Ref. [5] this yields our thickness dependent prediction for the rate of secondary bremsstrahlung following a primary photoabsorption event. To get a complete prediction we add the rates for these two processes (LET from Eqn. 2 and secondaries from Eqns. 6 and 7).

## C.    Results

Once all the data was combined and all corrections were made for each of the target thicknesses, we obtained our excess count $E$ given by $E \equiv C_o - A$ where $C_o$ is the number of coincidence counts and $A$ is the number of accidental counts averaged over 10 orbital periods. We found that the ratio of our uncertainties in excess counts to averaged accidental counts was in the range of 1% to 10%. Finally, we calculated as follows the quantity $W$, the fluorescence -normalized excess coincidence rate in terms of $E$ the excess counts, $F$ the fluorescence counts for the same run, $\langle \Omega_S \rangle$ the scattered photon detector solid angle averaged for the different choices of the scattered photon detector and $C$ (= 0.62) the average of the factor $\left(\frac{4}{5} - \frac{2}{5}cos^2\theta\right)$ in Eqn. (2) again averaged over scattered photon detector choices, $W \equiv E/(F \cdot C \cdot \langle \Omega_S \rangle)$.

As shown in Figure 7, we plotted $W$ for the different target thicknesses against scattered photon energy, $E_S$, along with the LET theory prediction given by Eqn. (2) and our secondary event calculations due to Eqn. (6) and Eqn. (7) as well as their sum, the expected total excess rate. We note that both theory and experiment are extremely weak effects: any excess signal is five orders of magnitude below the fluorescence signal. We see that at the lowest value of the scattered energy (2.5 keV) the measured $W$ values depart so strikingly, e.g., the $W$ value is significantly negative, an unphysical result, from the remaining spectra at higher energies that we regard them as unreliable for the purposes of discussion.

Following Ref. [5], we expect that the statistically significant signals seen with the 40, 160 and 320 nm targets near 8 keV are due to either a double fluorescence events following excitation of both K shell electrons or the electron impact ionization of a second atom by the photoelectron produced in the primary photoabsorption event. For the much thicker samples (7.7 microns) used in Ref. [5] such a signal could plausibly be due entirely to the later process as argued in Ref. [5]. For the considerably thinner targets used in the present work, a secondary scattering simulation is needed to resolve the matter. From the work of Briand *et al.*, Ref. [29], we expect that the presence of a double K vacancy would be indicated by an 8350 eV hypersatellite peak, but our 500 eV resolution is too poor to distinguish such a feature. Regardless, the emitted photons associated with double vacancy fluorescence or secondary processes are well separated from the expected regime for IAB.

Therefore, we now turn to the energy range of interest for the test of IAB theory: 3 to 7 keV as shown in Fig. 8. We note that by providing three successive doublings of the sample thickness we are scanning over a wide range of contribution of secondary scattering to the signal. We see that in the thinnest samples (40 nm.), the predicted secondary event rate based on Eqns. (6) and (7) is much smaller than the theoretical IAB signal while in the thickest sample they are about the same. Fig. 8 shows no sign of an excess signal for 3 to 7 keV for the 40, 80, and 160 nm. thick targets. However we see excellent evidence in the thickest targets (320 nm.) for a nonzero excess signal, roughly comparable to the expected secondary rate. To quantify the comparison between prediction and our observations Table II provides the $\chi^2$ values for comparisons of the observations in Figure 8 with various predictions. The right tailed probability[30] of consistency between predictions and the observations is based on using the number of data points, nine, as the number of degrees of freedom. We see that for the thinnest targets, 40 nm, where we expect the secondary signal to have a much lower strength than the LET prediction, there is some chance of the observations being consistent with no excess signal or the expected secondary signal alone, but the LET prediction added to the expected secondary is definitively ruled out[31] at the equivalent of a 5.4 sigma level.[32] This is the main result of this paper. For the next thickest sets of targets, 80 nm, all three possibilities are quite possible. The measurements for the 160 nm thick targets are quite consistent with the secondaries only or no excess but the addition of the expected LET signal to the secondary signal is clearly disfavored. For the thickest target where the secondaries are expected to be comparable to the LET signal and the uncertainties in the measurements are the very least, all models are inconsistent with the measurements, but as we noted the secondary signal emerges.

In summary, the LET prediction is decisively ruled out, the strongest evidence coming from the thinnest samples, where secondaries are not significant. From our thickest target data, we see the emergence of the secondary signals indicative of extrinsic Bremsstrahlung but our prediction is in need of refinement in order to go beyond semiquantitative agreement. Nevertheless the observation of this later signal at the expected level decisively validates our measurement and data reduction techniques. Thus, we have acquired sufficient statistics, dealt with extraneous backgrounds e.g. wings of the incident beam, possible extraneous interfering signals and properly normalized the crucial excess signal. Given that secondaries can only add to the observed excess signal beyond the expected IAB contribution, the severe upper limit we have placed on the later is reinforced. Pushing further, thanks to one of referees, the conclusion that the LET is exclude can be made quantitative independent of any model for the secondaries as follows: computing the chisquared for 40 nm samples for LET alone gives 39.1 with a consequent right tail probability of 1.1E-5. This implies a z-score of 4.2 sigma, again extremely unlikely. To summarize, through the use of ultrathin targets we have overcome the secondary scattering that has stymied progress in the experimental pursuit of a definitive test of the current IAB prediction.

## IV. CONCLUSIONS

The predicted spectra of low-energy photons ejected during beta decay and K capture events show agreement with experiment while the analogous spectra resulting from electron ejection upon x-ray photoabsorption generally do not. We provide a new measurement of the latter observationally contentious process to test the predictions that describe the infrared divergence of radiation accompanying the ejection of an electron from the inner shell of an atom. To do this we employed a contemporary fluorescence coincidence detection method to measure the rate of low energy inelastic photon scattering from the K-shell of copper targets of varying thickness exposed to x-ray photons from a synchrotron source. After normalizing by our observed fluorescence rate and making necessary corrections for absorption and accidental scattering events, we rule out the signal that is predicted by contemporary theory. Clearly, in contrast to the nuclear internal bremsstrahlung effect, we do not understand the intraatomic bremsstrahlung process.

## V. ACKNOWLEDGMENTS

This work is based upon research conducted at CHESS which was supported by the National Science Foundation under award DMR-1332208. This work made use of the Cornell Center for Materials Research Shared Facilities which are supported through the NSF MRSEC program under award DMR-1719875. We appreciate the aid and advice we received from Joel Brock, Ernest Fontes, Johnathan Kuan, Darren Pagan, Zachary Brown, James Shanks, Scott Smith, Jacob Ruff, Nate Rider, Sol Gruner, Elke Arenholz, Songqi Jia, Yue Li, Sagnik Saha, Vincent Marchetti and CHESS operators at Cornell, Ann Xiang, Christopher Cox and Peter Grudberg at XIA Corporation and our referees.

.

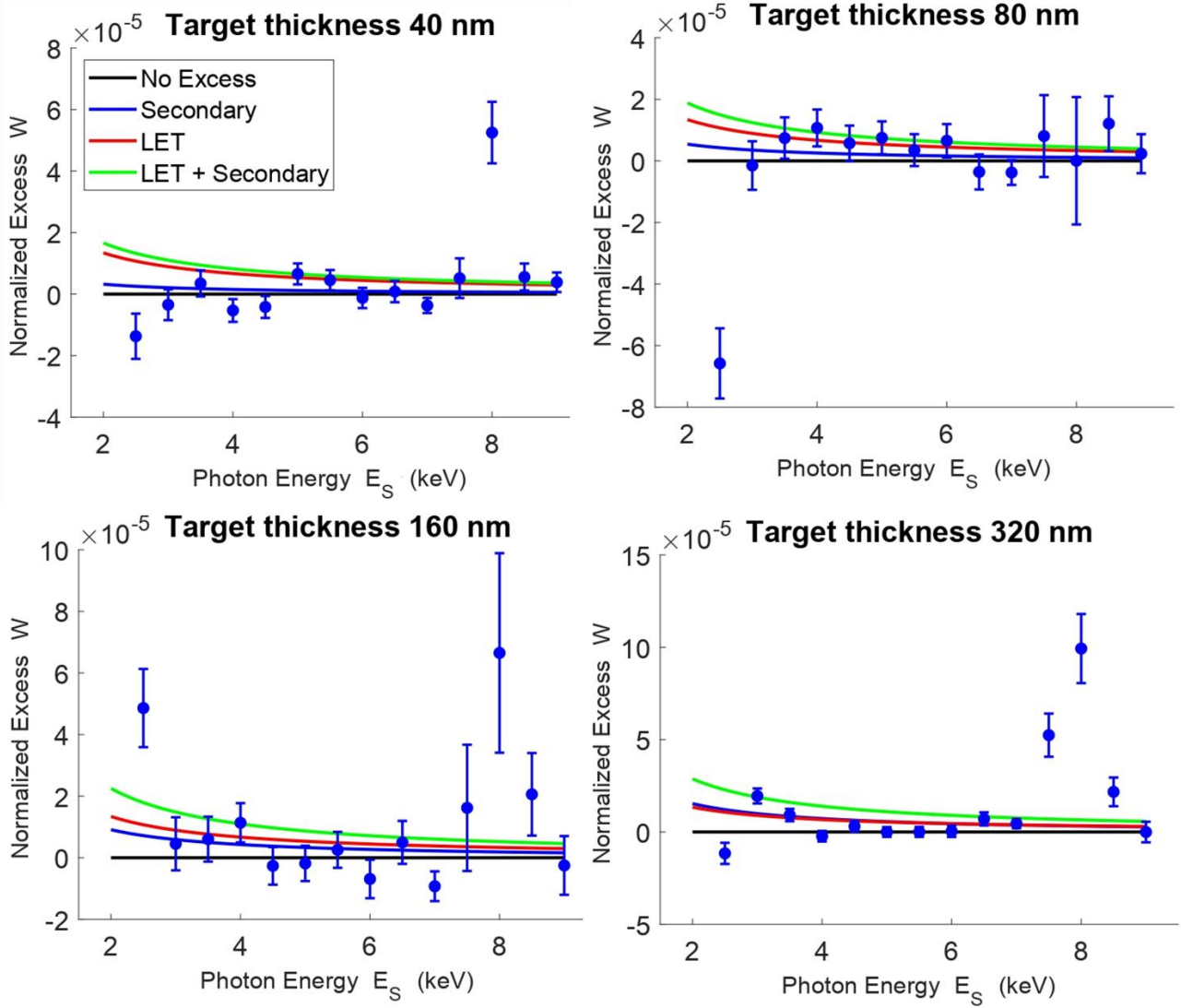

FIG. 7: Experimental values for $W$, Fluorescence-Normalized Excess Coincidence Rate Plotted Against $E_S$, Scattered Photon Energy for Each Sample Thickness. The expected results based on the LET prediction and secondary event calculation and their sum are given. For the thickest target there is a nonzero excess signal roughly in agreement with the expected secondary rate for the lowest energy. Possible sources of a signal around 8 keV are discussed in the main text. The order of curves with increasing W values are as follows: for 40, 80, 160 nm: no excess, secondary, LET, LET + secondary; for 320 nm: no excess, LET, secondary, LET + secondary.

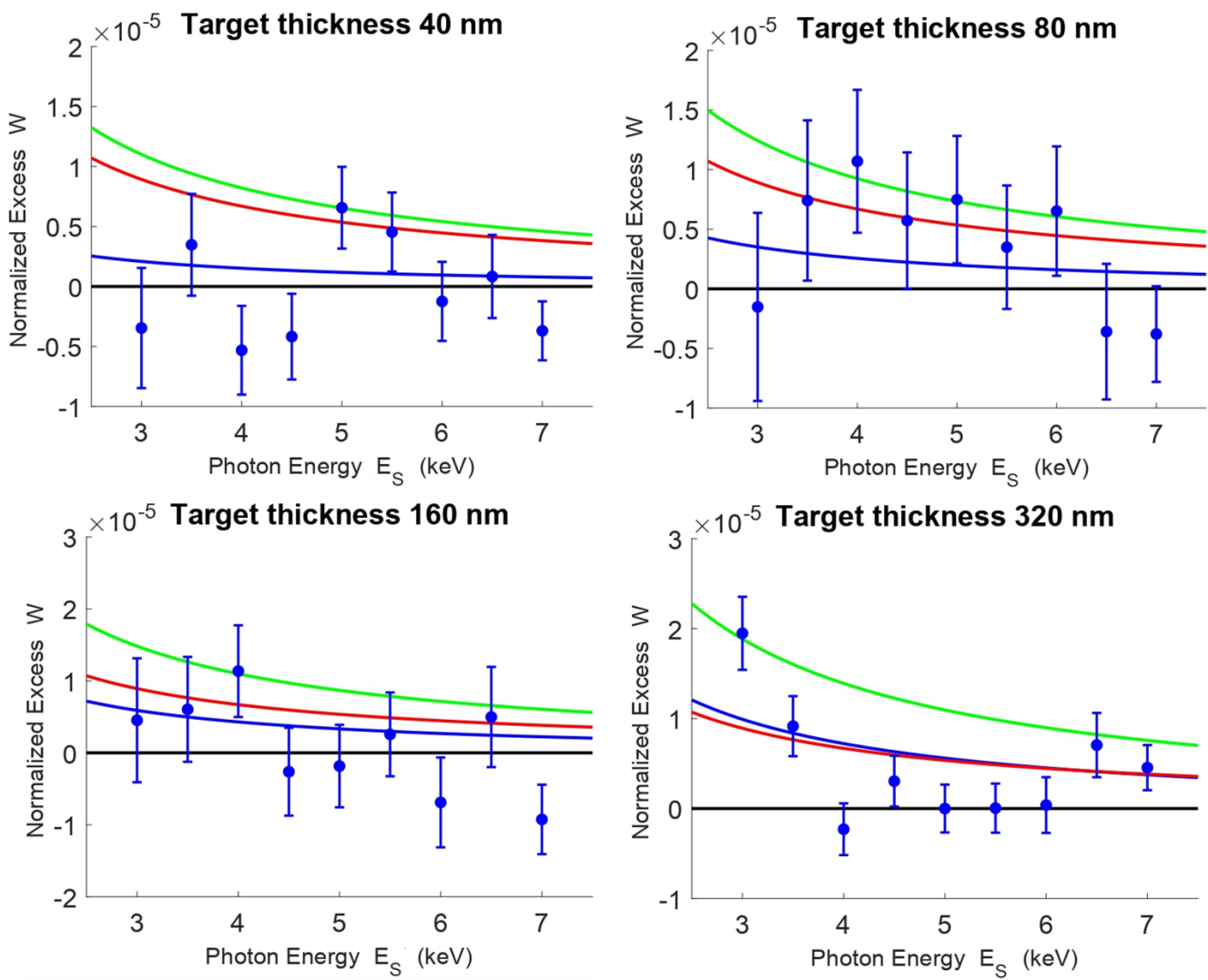

FIG. 8: Excess Coincidence Rate over the Scattered Photon Energy Range of Interest 3-7 keV. (Same plots as Figure 7 over this restricted energy range). The results for the thinnest sample (40 nm) strikingly rule out the LET prediction.

Table II: $\chi^2$ values in blue and Corresponding Right Tail Probabilities (RTP) of Observations in Figure 8 for Different Predictions Vs. Target Thickness,

| Thickness (nm) | LET and Secondaries | RTP | Secondaries alone | RTP | No excess | RTP |
|---|---|---|---|---|---|---|
| 40 | 51.1 | 6.6E-08 | 14.6 | 0.10 | 12.7 | 0.18 |
| 80 | 11.5 | 0.24 | 7.5 | 0.59 | 10.6 | 0.31 |
| 160 | 25.7 | 2.3-03 | 11.3 | 0.25 | 10 | 0.35 |
| 320 | 85.9 | 1.1E-14 | 28.2 | 8.8E-4 | 39.6 | 9.0E-6 |